\documentclass[letterpaper,titlepage,11pt]{article}

\usepackage{hyperref}
\usepackage{epsfig}
\usepackage{graphicx}
\usepackage{amsmath}
\usepackage{amsfonts}
\usepackage{amssymb}
\usepackage{mathrsfs}
\usepackage{fancybox}
\usepackage{ragged2e}
\usepackage{tikz}
\usepackage{wasysym}

\long\def\symbolfootnote[#1]#2{\begingroup%
\def\thefootnote{\fnsymbol{footnote}}\footnote[#1]{#2}\endgroup}

\newcommand{\mycite}[1]{~{\cite{#1}}}

\newcommand{\myref}[1]{~{(\ref{#1})}}

\newcommand{\beq}{\begin{equation}}
\newcommand{\eeq}{\end{equation}}
\newcommand{\be}{\begin{equation*}}
\newcommand{\ee}{\end{equation*}}

\numberwithin{equation}{section}

\begin{document}

\begin{titlepage}

\rightline{\vbox{\small\hbox{\tt } }}
 \vskip 1.8 cm
\centerline{\LARGE \bf Three-BMN Correlation Functions:} \vskip
0.3cm \centerline{\LARGE \bf Integrability vs. String Field Theory
One-Loop Mismatch} \vskip 1.cm
\centerline{\large  {\bf Waldemar Schulgin$\,^{1}$},  {\bf
A.~V.~Zayakin$\,^{2,3}$}}
\vskip 1.0cm
\begin{center}
\sl $^1$Universit\'e Libre de Bruxelles and International  Solvay
Institutes, ULB-Campus Plaine CP231, B-1050 Brussels, Belgium
\\
\vskip 0.4cm $^2 $ \sl Departamento de Fi\'sica de Part\'iculas\\
Universidade de Santiago de Compostela\\ and\\ Instituto Galego de
F\'isica de Altas Enerx\'ias (IGFAE)\\
E-15782 Santiago de Compostela, Spain\\
\vskip 0.4cm
\sl $^3$ Institute of Theoretical and  Experimental Physics,\\
B.~Cheremushkinskaya ul. 25, 117259 Moscow, Russia
\end{center}
\vskip 0.6cm

\centerline{\small\tt waldemar.schulgin@ulb.ac.be, a.zayakin@gmail.com }

\vskip 1.3cm \centerline{\bf Abstract} \vskip 0.2cm \noindent

We compare calculations of the three-point correlation functions
of BMN operators at the one-loop (next-to-leading) order in the
scalar SU(2) sector from the integrability expression recently
suggested by Gromov and Vieira, and from the string field theory
expression based on the effective interaction vertex by Dobashi
and Yoneya. A disagreement is found between the form-factors of
the correlation functions in the one-loop contributions.  The
order-of-limits problem is suggested  as a possible explanation of
this discrepancy.
\end{titlepage}

\tableofcontents
\section{Introduction}
Search for an exact matching between the perturbation theory
calculations of anomalous dimensions in the $\mathcal{N}=4$
supersymmetric Yang-Mills and string theory has been driving the
research in the AdS/CFT correspondence for a long time. It was
pointed out by Beisert\mycite{Beisert:2004ry} that the correct
comparison of the string and field-theoretical results would take
place only when the full non-perturbative expressions are being
compared. If one compares any expansions up to a certain degree
the comparison may be obstructed by the non-commutativity of the
limits. Namely, the string theory naturally admits the
thermodynamic limit as its basic assumption and then is decomposed
perturbatively in the coupling constant, whereas the field theory
intrinsically relies on the coupling constant perturbative
decomposition, while the thermodynamic limit is taken afterwards.
Thus already in\mycite{Janik:2006dc} Janik argued that a
discrepancy between the string theory  and the field theory may be
explained in terms of the order-of-limits problem. In the
two-point sector however the order-of-limits argument has finally
been found redundant, since the originally observed three-loop
discrepancy\mycite{Callan:2004uv} and the breakdown of the BMN
scaling at four loops\mycite{Serban:2004jf} was later cured not by
the invocation of the order-of-limits considerations but by the
introduction of the correct crossing-symmetric phase
factor\mycite{Beisert:2006zy,Beisert:2006ez} into the S-matrix.

Thus the order-of-limits argumentation, after having been
developed for explaining various discrepancies between the
anomalous dimensions on the weak and strong coupling sides, has
made place for more physical arguments instead. Now  that one is
in the possession of the full Bethe Ansatz for any coupling value
and any chain length\mycite{Beisert:2010jr}, the  anomalous
dimension of any operator is effectively known at either weak or
strong coupling at arbitrary precision.

The three-point functions in the $\mathcal{N}=4$ SYM present a new
challenge to the AdS/CFT correspondence statement. It has been
pointed out by Georgiou\mycite{Georgiou:2011qk} that even when an
agreement is observed for the structure constants $C_{123}$ at a
certain loop order, the agreement should fail at a higher power of
the coupling because of the order-of-limits problem. Thus the
strong vs weak coupling match or mismatch would be reduced to an
issue of a lucky coincidence, and would be devoid of physical
meaning. For example, a discrepancy between subleading orders in
$\lambda'$ expansion of the weak and strong coupling limits for a
heavy-heavy-light three-point correlator of scalars was reported
by Bissi, Harmark and Orselli\mycite{Bissi:2011ha}. The natural
question was how to interpret this result. On one hand, the
argumentation proposed by Harmark, Kristjansson and
Orselli\mycite{Harmark:2008gm} claimed that the near-BPS-states in
fact must necessarily comply with the string results up to
one-loop level. On the other hand, it has been stated
in\mycite{Escobedo:2011xw} that ``from a more modern perspective''
 the match of the spectra in the Frolov-Tseytlin limit be ``a
fortunate accident''.

We believe that the issue on whether the matching between the
structure coefficients in weakly or strongly coupled sector is an
accident still remains a valid question. The whole story of how
our knowledge of the spectra (i.e. the anomalous dimensions) of
the two-point functions developed is instructive for having
eliminated possible formal causes for different discrepancies in
exchanging them for a better understanding of the physics behind
the integrable chain both at the strong and at the weak coupling
limit. In particular, these were the discrepancies between the
perturbation theory and semiclassics that drove the discovery of
e.g. the dressing phase and the Y-system technique that endow us
with the full knowledge of the anomalous dimensions at any
coupling and any length. Therefore, we believe that it is of
utmost importance to collect the ``experimental evidence'' for
(mis)match of the weak and strong coupling results in the various
sectors of the theory even before we can interpret this (mis)match
properly, as already implemented for the $SO(6)$ sector in
comparing the direct perturbative calculation vs. string field
theory \mycite{Grignani:2012ur,Grignani:2012yu}, and for the
$SO(6)$-extended conjectured version of the Gromov-Vieira formula
against the string field theory in the leading
order\mycite{Bissi:2012vx}.

Here a test is performed at the next-to-leading order level for
the correlation functions of three $SU(2)$ BMN operators in the
weak and strong coupling limits. On the strong coupling side in
Section~\ref{sft} our basic approach is the string field theory
effective vertex quantum mechanics, using the approach suggested
originally in\mycite{Spradlin:2002ar,Spradlin:2002rv,He:2002zu}
and employing the correct prefactor for the effective vertex found
in\mycite{Dobashi:2004nm}. On the weak coupling side treated in
Section~\ref{gv} the basic technique is the integrability-assisted
calculation suggested by Gromov and Vieira\mycite{Gromov:2012uv}.
What is found as a result of our comparison is the disagreement
between the two calculations in the next-to-leading order, that
is, in the first order in $\lambda'$. We speculate in the
Conclusion on whether the possible physical causes of it should be
sought or this mismatch may be considered as a formal artifact
related to the order-of-limits problem.


\section{String Field Theory Computation\label{sft}}
The correlator of the three operators
$\mathcal{O}_1(x_1)\mathcal{O}_2(x_2) \mathcal{O}_3(x_3)$ is
characterized by its structure constant $C_{123}$ defined as
\beq \langle \bar{\mathcal{O}}_2(x_2) \mathcal{O}_1(x_1)
\mathcal{O}_3(x_3)\rangle=\frac{C_{123}}
{|x_1-x_2|^{\frac{\Delta_1+\Delta_2-\Delta_3}{2}}
|x_2-x_3|^{\frac{\Delta_2+\Delta_3-\Delta_1}{2}}
|x_3-x_1|^{\frac{\Delta_1+\Delta_2-\Delta_3}{2}}}, \eeq
where $\Delta_i$ are the dimensions of the operators. In this
section the three-point structure constant is calculated from the
point of view of string field theory. We follow the recipes
of\mycite{Dobashi:2004ka}. Namely, we start with their expression
(2.1) \begin{equation}\label{normalize}
C_{123}=\frac{1}{\mu(\Delta_1+\Delta_3-\Delta_2)} \left(f\frac{J_1
J_3}{J_2} \right)^{-\frac{\Delta_1+\Delta_3-\Delta_2}{2}}
\Gamma\left(\frac{\Delta_1+\Delta_3-\Delta_2}{2}+1
\right)\frac{\sqrt{J_1J_2J_3}}{N}\langle 1|\langle 2|\langle
3|H_3\rangle,\end{equation}
where \be \label{ff} f=\frac{1}{4\pi \mu r (1-r)}, \ee $J_i$ being
the R-charges of the respective chains, $\Delta_i$ the full
dimensions of the corresponding operators, $\langle 1|\langle
2|\langle 3|H_3\rangle$ the matrix element of the string effective
Hamiltonian. The parameter $\mu$ is related to the
Frolov--Tseytlin coupling $\lambda'$ as
\beq \mu=\frac{1}{\sqrt{\lambda'}.} \eeq

The concrete form  of the normalization factor on the right hand
side of (\ref{normalize}) was figured out in\cite{Dobashi:2004nm}
by expanding the result of an
integral of three bulk-to-boundary propagators in the strong
coupling regime for large $\Delta_i$ and neglecting all subleading
terms. For the holographic string theory dictionary this means
that the combinations of $\Delta_i$'s in this has to be taken only
at the leading order in $1/\mu$ and must not be expanded  further.
The subleading terms in $1/\mu$ will come only from the expansion
of the  matrix element on right and side of (\ref{normalize}).

The string field theory calculation has the property of yielding
always the finite result. Field-theoretically we interpret it as a
cancellation of the log divergences of two-particle external-leg
normalization with the proper three-particle divergences. Thus the
string field theory assumes that our basis is indeed the proper
basis of eigenstates in the respective order. It is well known
that the extremal correlators require the basis redefinition
already in the leading  ${\mathcal{O}}(1/N)$-order. The correlator
is said to be extremal if for the lengths $L_1,L_2,L_3$ of its
operators holds
\beq L_i+L_j-L_k=0. \eeq
Unlike those, the non-extremal correlators (for which
$L_i+L_j-L_k>0$ is always true) feel the basis redefinition only
for the subleading corrections. Happily enough, string field
theory based on the Dobashi--Yoneya improved vertex knows already
about these redefinitions\mycite{Dobashi:2004ka} and is therefore
applicable even to the extremal case. The use of the
Dobashi--Yoneya vertex and not of its earlier suggested analogs is
justified by the next-leading-order two-point
calculation\mycite{Grignani:2006en} that has been proven to be the
only vertex to yield the correct two-point subleading correlator.

The string field theory we are interested in is limited to the
``tree-level'' (leading topology,
$\mathcal{O}\left(\frac{J^2}{N}\right)$-order) contribution, thus
no string diagrams of the type considered in
e.g.\mycite{Grignani:2005yv} need to be considered. They certainly
do exist, but from\mycite{Grignani:2005yv} it is clear that the
loop string field theory effects are $1/N$ suppressed. Thus the
non-extremal correlator is a very neat object to be analyzed: if
the general framework of duality is correct, the tree-level result
in SFT is exact to all loops in terms of the gauge theory. The
gauge theory result is meant at weak coupling $g_{YM}$ and small
$\lambda^\prime$, whereas the string theory at $g_s=g^2_{YM}>>1$,
yet it is also taken at small $\lambda^\prime$, which allows the
comparison to be performed. To obtain the string field theory
result at the given loop order one needs to expand expression
(\ref{normalize}) up to the corresponding order in $1/\mu$.

The string vertex is organized as
\be |H\rangle = \sum_{m=0}^\infty\sum^3_{r=1}
\frac{\omega_m^{(r)}}{a_{(r)}} a^{(r)\dagger}_m a^{(r)}_m
|E\rangle,\ee
where the operators $a^{(r)\dagger}_m$ and $a^{(r)}_m$ are
creation and annihilation operators for the oscillator modes with
momentum number $m$, numeric coefficients\footnote{One should be
careful not to confuse the numbers $a_{(i)}$ with the creation and
annihilation operators. We stick to the notation of Dobashi and
Yoneya in \cite{Dobashi:2004ka}.}
 are
$a_{(1)}=r,a_{(2)}=-1,a_{(3)}=1-r$, the frequency is
$\omega_m^{(r)}=\sqrt{m^2+(\mu a_{(r)})^2}$, and the exponential
factor $|E\rangle$ looks like
\be |E\rangle = \exp\left[
-\frac{1}{2}\sum_{m,n=-\infty}^\infty\sum^3_{r=1}
\alpha_m^{(r)\dagger}\tilde{N}_{mn}^{rs}
\alpha_n^{(r)\dagger}\right]|0\rangle \, . \ee Notice the two
different bases of creation and annihilation operators used in the
same formula, related as
\be\begin{array}{l}  \displaystyle
\alpha_n=\frac{a_n-ia_{-n}}{\sqrt{2}},\\ \\\displaystyle
\alpha_{-n}=\frac{a_n+ia_{-n}}{\sqrt{2}}.
\end{array} \ee
Out of these two bases it is the $\alpha_n$ oscillators that
correspond directly to the BMN operators.
The $\tilde{N}$ matrices are taken from the work\mycite{He:2002zu}
(all indices $m,n$ assumed to be positive):
\be
\begin{array}{l}\displaystyle
\tilde{N}_{m,n}^{r,s}= \tilde{N}_{-m,-n}^{r,s}=
\frac{\bar{N}_{-m,-n}^{r,s}-\bar{N}_{m,n}^{r,s}}{2},\\
\displaystyle \tilde{N}_{m,-n}^{r,s}= \tilde{N}_{-m,n}^{r,s}=
-\frac{\bar{N}_{-m,-n}^{r,s}+\bar{N}_{m,n}^{r,s}}{2},
\end{array}
\ee
where the matrices $\bar{N}_{m,n}$ are
\be\begin{array}{l} \displaystyle
\bar{N}_{m,n}^{r,s}=\frac{1}{2\pi}
\frac{(-1)^{r(m+1)+s(n+1)}}{a_{(s})\omega_m^{(r)}+a_{(r)}\omega_n^{(s)}}
\sqrt{\frac{|a_{(s)}a_{(r)}|(\omega_m^{(r)}+\mu
a_{(r)})(\omega_n^{(s)}+\mu
a_{(s)})}{\omega_m^{(r)}\omega_n^{(s)}}},\\ \\ \displaystyle
\bar{N}_{-m,-n}^{r,s}=-\frac{1}{2\pi}
\frac{(-1)^{r(m+1)+s(n+1)}}{a_{(s)}\omega_m^{(r)}+a_{(r)}\omega_n^{(s)}}
\sqrt{\frac{|a_{(s)}a_{(r)}|(\omega_m^{(r)}-\mu
a_{(r)})(\omega_n^{(s)}-\mu
a_{(s)})}{\omega_m^{(r)}\omega_n^{(s)}}}.
\end{array}\ee
These definitions are $1/\mu$ exact up to any perturbative order:
only exponentially small corrections $\sim e^{-\mu}$ could be
absent from them.
For the non-extremal 3-BMN case one has
\be L_1=Jr+4, \quad L_2=J+2,\quad L_3=J(1-r)+2, \ee
and
\be \Delta_1=L_1+\frac{n_1^2+n_4^2}{\mu^2 r^2}
,\quad \Delta_2=L_2+\frac{n_2^2}{\mu^2 } ,\quad \Delta_1=L_3+\frac{
n_3^2}{\mu^2 (1-r)^2}, \ee
For the extremal 2-BMN case there is a pair of oscillators less,
thus \be L_1=Jr+2, \quad L_2=J+2,\quad L_3=J(1-r), \ee
and
\be \Delta_1=L_1+\frac{n_1^2}{\mu^2 r^2} , \quad
\Delta_2=L_2+\frac{n_2^2}{\mu^2 },\quad \Delta_3=L_3. \ee
To calculate the matrix element $\langle 1|\langle 2|\langle
3|H_3\rangle$ for the three BMN case all the possible contractions
are considered between the four magnons with the momenta
$n_1,-n_1,n_4,-n_4$ and the other four magnons with momenta
$n_2,-n_2,n_3,-n_3$. There are 24 such contractions of the type
\beq F_{abcd,a'b'c' d'}\equiv \tilde{N}^{12}_{a a'}
\tilde{N}^{12}_{bb'} \tilde{N}^{13}_{cc'} \tilde{N}^{13}_{dd'},
\eeq
where $a,b,c,d$ take the values of $n_1,-n_1,n_4,-n_4$,
$a',b',c',d'$ those of $n_2,-n_2,n_3,-n_3$ in all possible
combinations. It should also be taken into account that the
prefactor written in terms of $\alpha^{i\dagger}_m,\alpha^{i}_m$
operators looks like
\beq P\equiv \sum_n\sum^3_{r=1} \frac{\omega_n^{(r)}}{a_{(r)}}
\left(\alpha^{(r)\dagger}_n a^{(r)}_{n}+\alpha^{(r)\dagger}_{-n}
a^{(r)}_{-n}+\alpha^{(r)\dagger}_n
a^{(r)}_{-n}+\alpha^{(r)\dagger}_{-n} a^{(r)}_n\right) .\eeq
Therefore, while contracting the matrix element
\be \langle 1|\langle 2|\langle 3|H_3\rangle=\langle
\alpha^{(1)}_{n_1}\alpha^{(1)}_{-n_1}
\alpha^{(1)}_{n_4}\alpha^{(1)}_{-n_4}
\alpha^{(2)}_{n_2}\alpha^{(2)}_{-n_2}
\alpha^{(3)}_{n_3}\alpha^{(3)}_{-n_3} |P|E\rangle \ee
some of the momenta ``change their sign'' when they are contracted
with $|E\rangle$  through the prefactor $P$. Thus denoting
auxiliary quantity
\be
\begin{array}{l}
F^{(1)}_{n_1,n_2,n_3,n_4}= F_{n_1,n_2,n_3,n_4,-n_1,-n_2,-n_3,-n_4}
+F_{-n_1,n_2,n_3,n_4,n_1,-n_2,-n_3,-n_4}+\\
+F_{n_1,n_2,n_3,n_4,n_1,-n_2,-n_3,-n_4}+
F_{-n_1,n_2,n_3,n_4,-n_1,-n_2,-n_3,-n_4},
\end{array}
\ee
and similarly defining
$F^{(2)}_{\dots},F^{(3)}_{\dots},F^{(4)}_{\dots}$ for the
permutations of the signs of magnon momenta $n_2,n_3,n_4$
respectively one obtains finally the matrix element
\be
\langle 1|\langle 2|\langle 3|H_3\rangle=\sum_{n_i\in
{\rm{magnons}}}\frac{\omega_{n_i}^{(r)}}{a_{(r)}}
F^{(i)}_{n_1,n_2,n_3,n_4} \, .
\ee
Combinatorics for the two-BMN case is derived analogously. Taking
these matrix elements $\langle 1|\langle 2|\langle 3|H_3\rangle$
for the two-BMN and three-BMN cases together with the
normalization factors of\myref{normalize},  the following results
are obtained on the string field theory side.

{\bf (A)} For the three-BMN case we obtain
\begin{eqnarray}\displaystyle
C_{SFT,3BMN}|_{n_4\to n_1}=\frac{1}{N \sqrt{J}} \frac{16\sqrt{r}
\left(3 n_2^2 r^2+n_1^2\right) \sin ^2\left(\pi n_2
r\right)}{\pi ^2\sqrt{(1-r)} \left(n_1^2-n_2^2 r^2\right){}^2}
\times \nonumber\\ \nonumber  \\
\displaystyle  \hphantom{=====} \left[1+ \frac{\lambda'}{4}
\left(-\frac{3 n_1^2}{r^2}-2n_2^2 -\frac{2 n_3^2}{(r-1)^2}+
\frac{12 n_1^2n_2^2}{3 n_2^2 r^2+n_1^2}\right)\right] \,
.\label{SFT3}
\end{eqnarray}
As has been mentioned above, the $1/\mu$-expanded expression for
the correlator inherits the $\mu$-dependence from the Neumann
matrices and the effective vertex prefactor $P$, yet not from the
$\Delta_i$ in the normalization factor
$\frac{1}{\mu(\Delta_1+\Delta_3-\Delta_2)} \left(f\frac{J_1
J_3}{J_2} \right)^{-\frac{\Delta_1+\Delta_3-\Delta_2}{2}}
\Gamma\left(\frac{\Delta_1+\Delta_3-\Delta_2}{2}+1 \right)$ by
Dobashi-Yoneya. There has been no weak-coupling calculation for
the three-BMN correlator so far, thus it will be compared to the
weakly-coupled side after the three-point correlator is computed
using integrability in the next section.

{\bf (B)} For the extremal correlator of two-BMN one-BPS one
obtains
\beq \label{extrSFT}\displaystyle C_{SFT,2BMN}= \frac{2 J^{3/2}}{N}
\frac{\sqrt{r^7(1-r)}n_2^2 \sin ^2\left(\pi n_2 r\right)}{\pi ^2
\left(n_1^2-n_2^2 r^2\right){}^2} \left[1+
\frac{\lambda'}{4} \left(\frac{n_1^2}{r^2}-n_2^2\right)\right] \,
.\eeq
Comparing this to the result of\mycite{Beisert:2002bb} for the
extremal correlator
\beq \label{extrFT}C_{FT,2BMN}= \frac{2 J^{3/2}}{N} \frac{\sqrt{r^7(1-r)}n_2^2
\sin ^2\left(\pi  n_2 r\right)}{\pi ^2 \left(n_1^2-n_2^2
r^2\right){}^2} \left[1+ \frac{\lambda'}{2}
\left(\frac{n_1^2}{r^2}-n_2^2\right)\right]\, , \eeq
one sees that it does not match at next-to-leading order.

For the three-point function to have a proper  scaling the
operators have to be the eigenstates of the dilatation operator.
In general, the eigenstates are  mixed states between single and
double trace operators. In the case of the extremal correlators
the contribution from the double-trace operators could be of the
same order in $1/N$ as from the single trace operators. The mixing
at ${\cal O}(\lambda')$-level affects the  ${\cal O}(\lambda'^0)$
and ${\cal O}(\lambda')$-contributions to the three-point
function. The one-loop contribution becomes also affected  from the
mixing at two-loop level.

Having this in mind we should note that the result of
\mycite{Beisert:2002bb} does include the mixing only at one-loop
level where the string field theory computation should capture
the mixing also at two-loop level. This makes the discrepancy at
${\cal O}(\lambda')$ in (\ref{extrSFT}) and (\ref{extrFT})
plausible.


\section{Integrability-assisted computation\label{gv}}
The direct perturbative calculation of three-point function is
straightforward and has been implemented since a long time. An
ambitious project to cast the calculation of the three-point
functions
 into the formalism of Bethe Ansatz was proposed
in\mycite{Escobedo:2010xs} and realized there at the leading order
in coupling constant. This ``three-point-functions from
integrability'' framework has been certainly inspired by the
success of integrable systems describing the two-point functions
(that is, the spectra of anomalous dimensions). In the leading
order integrability has provided a combinatorial simplification,
which is very important for calculating the correlation functions
with more than two excitations per operator. Yet it is the
next-to-leading order result of\mycite{Gromov:2012uv} that allows
one to fully appreciate the convenience of the Bethe Ansatz
calculation compared with the ordinary perturbation theory.
Physically, the integrability calculation does not yet provide us
with any new information like higher-order Hamiltonians or
semiclassical descriptions of highly-excited states. However, a
direct perturbative calculation would have required from us an
explicit knowledge of the interaction Hamiltonians and the
fudge-factors, the latter becoming increasingly more complicated
with larger numbers of magnons. Surprisingly, integrability
becomes a natural language to describe these complicated objects
in terms of scalar products of Bethe states; contributions of the
fudge-factors and Hamiltonian insertions are shown to be nicely
packed into a simple structure of a determinant of a matrix, the
size of which is proportional to the number of magnons rather than
e.g. to the operator length. This simplifies the problem
significantly and {\it de facto} proposes a new formalism rather
than a rewriting of an old one. Thus, although ``integrability
calculation'' is not independent physically from perturbation
theory, it exists at the present level as a very special formalism
that can be considered as a separate entry in the register of
duality recipes. Therefore the tests done on the weak side of
duality are expressly performed as tests of either perturbative
theory, or ``integrability-assisted'' perturbation theory.

As shown by Gromov and Vieira in \cite{Gromov:2012uv}
integrability allows us to build up the expression for the
three-point function structure coefficient up to the
${\mathcal{O}}(g^2)$-order out of the following ingredients
\begin{equation}
C_{{\rm 123}}=\rm{\texttt{norms}}\times \rm{\texttt{simple}}
\times\rm{\texttt{involved}}+{\mathcal{O}}(g^4)\ . \nonumber
\end{equation}
Below we discuss the form and the meaning of each of these
ingredients. Roughly speaking they can be understood in the
following way: the $\rm{\texttt{involved}}$ factor contains matrix
elements of the operator $\mathcal{O}_3$ between the states $1$
and $2$. The $\rm{\texttt{norms}}$ factor precisely corresponds to
the norms of Bethe states. The $\rm{\texttt{simple}}$ factor
represents a phase of the wave-function of the third operator that
is generated when transforming the rest of the expression into the
structure of the $\langle 1 |\mathcal{O}_3| 2\rangle$ structure.

To apply this formalism the operators should  be the Bethe
eigenstates at two-loop level.  The operator  lengths are denoted
as $L_1,L_2,L_3$ with corresponding number of magnons
$N_1$, $N_2$, $N_3$ and Bethe vectors
${\bf{u}}=\{u_i\},{\bf{v}}=\{v_i\},{\bf{w}}=\{w_i\}$. Below we
shall use the terms ``Bethe vector'', ``Bethe state'' and
``operator'' as complete synonyms.

\begin{itemize}
\item The first building block, \texttt{norms},  has the
form\footnote{In \cite{Gromov:2012uv} there are two conventions
which go by the names of algebraic and coordinate Bethe ansatz
normalizations, $ \langle {\bf u}|{\bf u}\rangle_{\bf
co}=\frac{1}{\mu^2}\langle {\bf u}|{\bf u}\rangle_{\bf  al} $ with
\begin{equation}
\mu=\left(1-g^2\Gamma_{\bf u}^2\right)
\prod_{j=1}^N\left(\frac{x(u_j-i/2)}{x(u_j+i/2)}-1\right)
\prod_{i<j}{\mathfrak f}(u_i,u_j) \ , \qquad
{\mathfrak{f}}=\left(1+\frac{i}{u-v}\right)
\left(1+\frac{g^2}{(u^2+\frac{1}{4})(v^2+\frac{1}{4})}\right)\nonumber
\end{equation}
In this work we use algebraic  Bethe ansatz normalization. }

\begin{equation}
\texttt{norms}=\frac{L_1 L_2 L_3}{\sqrt{\langle  {\bf{u}}
|{\bf{u}}\rangle\langle {\bf{v}} |{\bf{v}}\rangle\langle {\bf{w}}
|{\bf{w}}\rangle}} \ ,
\end{equation}
where
\begin{eqnarray}
\langle {\bf u}|{\bf u}\rangle&=&\left(1-2g^2
 \Gamma_{\bf u}-g^2 \Gamma^2_{\bf u}\right)
 \Big(\!\!\!\Big(\langle\theta ;{\bf{u}}|\theta
  ;{\bf{u}}\rangle\Big)\!\!\!\Big)_{\theta}\ , \nonumber\\
\Big(\!\!\!\Big(\langle\theta ;{\bf{u}}|\theta
 ;{\bf{u}}\rangle\Big)\!\!\!\Big)_{\theta}&=&\prod_{m\neq k}\frac{u_k-u_m+i}{u_k-u_m}\det_{j,k\leq N_1}\frac{\partial}{\partial u_j}\left(
\frac{L}{i}\log\frac{x(u_k+i/2)}{u_k-i/2}+\frac{1}{i}\sum_{m\neq k}^{N_1}\log\frac{u_k-u_m-i}{u_k-u_m+i}
\right) \, ,
\nonumber\\
\Gamma_{\bf{u}}&=&\sum_{i=1}^{N_i}\frac{1}{u_i^2+\frac{1}{4}}\ ,\qquad
x(u)=\frac{u+\sqrt{u^2-4 g^2}}{2g}=\frac{u}{g}-\frac{g}{u}+\ldots
\end{eqnarray}

\item The expression $\rm \texttt{simple}$ that corresponds to the
phases of the third operator's wave-function has the form

\begin{equation}
{\rm \texttt{simple}}= \left(1-g^2\Gamma_{{\bf w}}\right){\cal
A}_{N_3}(\bf{p}),
\end{equation}
where
\begin{eqnarray}
{\cal A}_N({\bf{p}})=
(1-g^2\Gamma^2_{\bf{w}}) \sum_{\alpha\cup \bar\alpha
=\{p\}}(-1)^{|\alpha|}\prod_{k\in\alpha,\bar k\in
\bar\alpha}\mathfrak{f}(k,\bar k)\prod_{\bar k\in
\bar\alpha}e^{iN\bar k}\nonumber\\
\end{eqnarray}
with $\alpha,\bar\alpha$ all possible partitions of the set of
momenta, and $|\alpha|$ number of the elements in partition
$\alpha$, and
\begin{equation}
{\mathfrak{f}}(k,\bar k)=\left(1+\frac{i}{w(k)- w(\bar
k)}\right)\left(1+\frac{g^2}{(w(k)^2+\frac{1}{4})(w(\bar
k)^2+\frac{1}{4})}\right)\, .
\end{equation}

To simplify the building blocks of the final expression  of the
three-point function we denote the expression for the
${\cal{A}}_N$ without the $(1-g^2\Gamma^2_{\bf{w}})$ prefactor
$\tilde{\cal{A}}_N$.\footnote{The relation derived in
\cite{Gromov:2012uv} is given in the coordinate Bethe ansatz
normalization. Note that the relation between ${\cal{A}}_N$ in
coordinate and algebraic Bethe ansatz normalizations is
$${\cal{A}}^{(\rm{alg})}_N=\mu {\cal{A}}^{(\rm{coord})}_N$$.}

\item The expression \texttt{involved} corresponds to  the scalar
product of Bethe states. Scalar products of Bethe states lie at
heart of the simplification offered by the integrability framework
for the three-point correlation functions. The scalar products for
Bethe eigenstates of  simplest groups (e.g. $SU(2)$) were built as
early as in the eighties (see
e.g\mycite{Reshetikhin:1983vw,resh,izergin1987,Slavnov1989}),
later this technique was extended towards more complicated cases
(non-compact groups, non-eigenstate vectors in the scalar
products); see also references in e.g.\mycite{Foda:2012wn}. The novelty
of~\cite{Gromov:2012uv} was to use the scalar product for
calculations of physical quantities, the three-point functions.

The scalar product could be expressed with the help  of the
so-called theta-morphism, which is a particular linear
transformation of a function $f$ that is related to some
homogeneous integrable chain of length $L$. Introduce
inhomogeneities $\theta_{i}, i=1\dots L$, one per chain node, into
the chain; then the theta-morphism $((f))_{\theta}$ of the
function $f$ is defined as
\begin{equation}
\Big(\!\!\!\Big(f(\theta)\Big)\!\!\!\Big )\equiv
f\Bigg|_{\theta_i\rightarrow 0}+\frac{g^2}{2}\sum_{i=1}^L{\cal
D}_i^2 f\Bigg|_{\theta_i\rightarrow 0}\ ,
\end{equation}
where
\begin{equation}
{\cal D}_i=\frac{\partial}{
\partial\theta_{i}}-\frac{\partial}{\partial\theta_{i+1}}, \qquad
{\rm{with}} \qquad \theta_{L+1}=\theta_1\nonumber \ ,
\end{equation}
see appendix \ref{thetamorphism} for more properties. Not going
deeply into the physical origin, derivation and the meaning of the
theta-morphism itself let us write out the recipe for the
``involved'' part of the calculation; in combination with the
norms of the first two operators it could be written as
\begin{equation}
\frac{{\rm involved}}{\sqrt{\langle  {\bf{u}}  |{\bf{u}}   \rangle
\langle {\bf{v}}  |{\bf{v}} \rangle}}= 
\frac{ \Big(\!\!\!\Big(\langle\theta^{(1)} ;{\bf{u}}|
\hat{\cal{O}}_3 |\theta^{(2)}
;{\bf{v}}\rangle\Big)\!\!\!\Big)_{\theta^{(1)}} }{\sqrt{
\Big(\!\!\!\Big(\langle\theta^{(1)} ;{\bf{u}}|\theta^{(1)}
;{\bf{u}}\rangle\Big)\!\!\!\Big)_{\theta^{(1)}}
\Big(\!\!\!\Big(\langle\theta^{(2)} ;{\bf{v}}|\theta^{(2)}
;{\bf{v}}\rangle\Big)\!\!\!\Big)_{\theta^{(2)}} }} + {\rm {pure\
imaginary \ term}}\, .
\end{equation}
All other building blocks of the final
expression  of the three-point function are real at ${\cal{O}}(g^2)$-order.
This allows to absorb the imaginary term into the overall complex
phase (the imaginary part being of order  ${\cal{O}}(g^2)$ could influence the
magnitude of the absolute value of the structure coefficient only
at the $g^4$-level which we neglect).

Scalar product involving  operators ${\bf 1}$ and ${\bf 2}$ is
given by
\begin{eqnarray}
\langle\theta^{(1)};{\bf u}|  \hat{\cal
O}_3|\theta^{(2)};{\bf{v}}\rangle= \frac{
{\displaystyle\prod_{m}^{N_3}}{\displaystyle\prod_{n}^{N_1}}(u_n-\hat
\theta_m^{(1)}+i/2)/{\displaystyle\prod_m^{N_3}}{\displaystyle\prod_n^{N_2}}(v_n-\hat\theta_m^{(1)}+i/2)
} { {\displaystyle \prod_{n< m}^{N_1}}(u_m-u_n){\displaystyle
\prod_{n< m}^{N_2}}(v_n-v_m){ \displaystyle\prod_{n<
m}^{N_3}}(\hat\theta_n^{(1)}-\hat\theta_m^{(1)})} \det\Big(
[G_{nm}]\oplus[F_{nm}]
\Big).\nonumber\\
\end{eqnarray}
The parameters  $\theta_m^{(r)}$ living on the nodes , where
$r=1,2,3$, $m=1\dots L_r$, are auxiliary quantities necessary to
perform the theta-morphism operation. Here where $\hat
\theta_m^{(1)}=\hat \theta_{L_1+1-m}^{(1)}$ and
\begin{equation}
F_{nm}=\frac{1}{\left(u_n-\hat \theta_m^{(1)}\right)^2+\frac{1}{4}}\ , \qquad G_{nm}={\displaystyle \prod_{a=1}^{L_1}}\frac{v_m-\theta_a^{(1)}+i/2}{v_m-\theta_a^{(1)}-i/2}\frac{\prod_{k\neq n}^{N_1}(u_k-v_m+i)}{u_n-v_m}-\frac{\prod_{k\neq n}^{N_1}(u_k-v_m-i)}{u_n-v_m}
\end{equation}
\end{itemize}
Combining previously discussed  pieces gives at  ${\cal{O}}(g^2)$-order up to a complex phase
factor
\begin{eqnarray}\label{GVIV}
C_{123}&=&\sqrt{L_1 L_2 L_3} \times\frac{\left(1+g^2 \Gamma_{\bf
w}+g^2 \Gamma^2_{\bf w}/2\right)}
{\sqrt{\Big(\!\!\!\Big(\langle\theta ;{\bf{w}}|\theta
;{\bf{w}}\rangle\Big)\!\!\!\Big)_{\theta}}}
\times{\rm{simple}}\times \frac{{\rm involved}}{\sqrt{\langle {\bf{1}}  |{\bf{1}}   \rangle \langle {\bf{2}}  |{\bf{2}} \rangle}}\nonumber\\
&=&\frac{ \sqrt{L_1 L_2 L_3}\Big(\!\!\!\Big(\langle\theta^{(1)}
;{\bf{u}}| \hat{\cal{O}}_3 |\theta^{(2)}
;{\bf{v}}\rangle\Big)\!\!\!\Big)_{\theta^{(1)}}\left(1-g^2\Gamma_{\bf{w}}/2\right)
\tilde{\cal{A}}_{N_3}({\bf{w}}) }{\sqrt{
\Big(\!\!\!\Big(\langle\theta^{(1)}  ;{\bf{u}}|\theta^{(1)}
;{\bf{u}}\rangle\Big)\!\!\!\Big)_{\theta^{(1)}}
\Big(\!\!\!\Big(\langle\theta^{(2)} ;{\bf{v}}|\theta^{(2)}
;{\bf{v}}\rangle\Big)\!\!\!\Big)_{\theta^{(2)}}
\Big(\!\!\!\Big(\langle\theta^{(3)} ;{\bf{w}}|\theta^{(3)}
;{\bf{w}}\rangle\Big)\!\!\!\Big)_{\theta^{(3)}} }}\, .
\end{eqnarray}

\subsection{BMN-BMN-BMN correlator}
In this section the above  formalism is applied to a specific
computation of a three point function of three BMN operators. We
consider a configuration of three BMN operators with lengths
\begin{equation}
L_1=r J+4\ , \qquad  L_2= J+2\ , \qquad L_3= J(1-r)+2\ ,
\end{equation}
the corresponding numbers of magnons are $N_1=4, N_1=2, N_3=2$ and
the rapidities are
\begin{eqnarray}
{\cal O}_1&: & u_1,-u_1,u_3,-u_3 \ ,\nonumber\\
{\cal O}_2&: & v,-v \ ,\nonumber\\
{\cal O}_3&: & w,-w \ .
\end{eqnarray}
Inserting all the ingredients into the expression  (\ref{GVIV})
(for details see Appendix \ref{details}) one gets
\begin{eqnarray}
C_{GV,3BMN}\Big|_{n_4\rightarrow n_1}&=&\frac{1}{
N\sqrt{J}}\frac{16\sqrt{r}\sin^2(\pi n_2 r)}{\sqrt{(1-r)}\pi^2}\
\frac{n_1^2+3n_2^2 r^2}{(n_1^2-n_2^2 r^2)^2} \left[
1+\frac{\lambda'}{4}\left(
-\frac{2n_3^2}{(1-r)^2}-\frac{n_1^2}{r^2}-2n_2^2+\frac{8n_1^2n_2^2}{n_1^2+3n_2^2r^2}
\right)
\right] \, ,\nonumber\\
\end{eqnarray}
where the limit  $n_4\rightarrow n_1$ has been taken at the very
end to keep the expression compact. We also used
$\lambda'=\frac{16\pi^2 g^2}{J^2}$. This matches the SFT
result\myref{SFT3} at the leading orders, but disagrees with it at
the subleading order.

\subsubsection{Comparison to the result of the string field
theory computation}

As demonstrated above there is complete matching of the three-BMN
correlators in the leading order. This term with $n_4$-dependency
restored has the following form
\begin{eqnarray}
C^0_{GV,3BMN}&=&C^0_{SFT,3BMN}=\nonumber\\
&=&\frac{1}{N\sqrt{J}}\frac{8 \sqrt{r} \sin ^2(\pi {n_2} r)}{\sqrt{ (1-r) }\pi ^2 }\frac{ \left(n_2^2 r^2 \left(n_1^4+n_4^4\right)-5
  n_2^4 r^4 \left(n_1^2+n_4^2\right)+n_1^2n_4^2
   \left(n_1^2+n_4^2\right)+6n_2^6 r^6\right)}{
   \left(n_1^2-n_2^2 r^2\right)^2 \left(n_4^2-n_2^2 r^2\right)^2}\, .\nonumber\\
\end{eqnarray}
The mismatch is happening at the ${\cal O}(\lambda')$-order. The
difference is coming from the terms depending on $n_1$ and $n_4$.
To illustrate the difference the $n_4$ is restored  in one of
these terms, which then takes the form
\begin{eqnarray}
C^0_{GV,3BMN}\left(
1+\frac{\lambda'}{4}\left(
\ldots-\frac{n_1^2}{r^2} +
\ldots
\right)
\right)&\longrightarrow& C^0_{GV,3BMN}\left(
1+\frac{\lambda'}{4}\left(
\ldots-\frac{1}{r^2}\frac{n_1^4+n_4^4}{n_1^2+n_4^2} +
\ldots
\right)
\right)\, ,\nonumber\\
C^0_{SFT,3BMN}\left(
1+\frac{\lambda'}{4}\left(
\ldots-\frac{3n_1^2}{r^2} +
\ldots
\right)
\right)&\longrightarrow& C^0_{SFT,3BMN}\left(
1+\frac{\lambda'}{4}\left(
\ldots-\frac{2}{r^2}\frac{n_1^4+n_1^2 n_4^2+n_4^4}{n_1^2+n_4^2} +
\ldots
\right)
\right)\, .\nonumber\\
\end{eqnarray}
The different structures between  these expressions will not allow
to match both expression by sending $n_1$ or $n_4$ to zero.
However, an interesting feature of this result is the observation
that sending both $n_1$ and $n_4$ at the same time to zero will
yield a matching expression on both sides
\begin{equation}
C_{3BMN}\Bigg|_{n_1,n_4\rightarrow 0}=
\frac{1}{
N\sqrt{J}}\frac{48\sin^2(\pi n_2 r)}{\sqrt{r^3(1-r)}\pi^2}\
\frac{1}{n_2^2 } \left(
1-\frac{\lambda'}{2}\left(
\frac{n_3^2}{(1-r)^2}+n_2^2
\right)
\right) \, .
\end{equation}


\subsection{BMN-BMN-BPS correlator}
In the formalism by Gromov--Vieira,  the case of the extremal
three-point correlators, like e.g. BMN-BMN-BPS, could be computed
via
\begin{equation}\label{extremal}
C_{123}=\frac{
\sqrt{L_1 L_2 L_3}\Big(\!\!\!\Big(\langle\theta^{(1)} ;{\bf{u}}| \hat{\cal{O}}_3 |\theta^{(2)} ;{\bf{v}}\rangle\Big)\!\!\!\Big)_{\theta^{(1)}}
}{\sqrt{
\Big(\!\!\!\Big(\langle\theta^{(1)} ;{\bf{u}}|\theta^{(1)} ;{\bf{u}}\rangle\Big)\!\!\!\Big)_{\theta^{(1)}}
\Big(\!\!\!\Big(\langle\theta^{(2)} ;{\bf{v}}|\theta^{(2)} ;{\bf{v}}\rangle\Big)\!\!\!\Big)_{\theta^{(2)}}
}}\, .
\end{equation}
However, there  is a subtlety involved concerning the mixing
between single and double trace operators. At one loop the
operators with well defined scaling dimensions are the mixed
states of the  single and double trace operators. In the
non-extremal case, like e.g. the three point correlation function
of three BMN operators, the contribution from the double trace
operators at ${\cal O}(g^2)$ order is always subleading in $1/ N$
compared to the one of the single trace operators. In the extremal
case, the contribution from the double trace operators is of the
same order in $1/N$ and becomes relevant already at the ${\cal
O}(\lambda'^0)$ order.

The formalism of Gromov/Vieira uses mappings  between the single
trace operators and the Bethe eigenstates which means this
formalism cannot give the complete three-point function but only
the contribution from the single trace operators. This phenomenon
could also be seen explicitly in the case  of the BMN-BMN-BPS
three point function. The eq. (\ref{extremal}) is applied to the
configuration with the lengths
\begin{equation*}
L_1=r J+2\ , \qquad  L_2= J+2\ , \qquad L_3= J(1-r)\ ,
\end{equation*}
and the rapidities
\begin{eqnarray}
{\cal O}_1&: & u,-u \ ,\nonumber\\
{\cal O}_2&: & v,-v \ , \nonumber
\end{eqnarray}
and obtain
\begin{equation}
C_{GV,2BMN}=  \frac{2 J^{3/2}}{N} \frac{\sqrt{r^7(1-r)}n_2^2 \sin
^2\left(\pi n_2 r\right)}{ \pi ^2 \left(n_1^2-n_2^2
r^2\right){}^2} \left[\left(1+\frac{n_1^2}{n_2^2r^2}\right)-
\frac{\lambda'}{2} \frac{(n_1^2-n_2^2 r^2)^2}{n^2_2 r^4}\right] \ .
\end{equation}
The mixing with the double  trace operators is already relevant at
the tree level of the three point function. Note that the extremal
three-point function at ${\cal O}(\lambda')$-order might also get
some contributions from the mixing of the operators at two-loop
order. As shown in \cite{Constable:2002vq},  the inclusion of the
double trace operators will change the tree level contribution by
a factor
\begin{equation}
C_{123}^{\rm{double\
trace}}=\left(1+\frac{n_1^2}{n_2^2r^2}\right)C_{123}^{\rm{without \
double \ trace}} \, .
\end{equation}
With the one-loop mixing  contribution taken into account it is
clear that the ${\cal O}(\lambda'^0)$ of the integrability
calculation matches the perturbative computations, see e.g.
\cite{Beisert:2002bb}.

Concerning the $\lambda'$-correction we are aware of two
perturbative computations in the above sector. The one by Beisert
et al. \cite{Beisert:2002bb} apparently takes into account the
contributions from the mixing of the single trace with double
trace operators and that's why cannot be compared. The authors of
\cite{Chu:2002pd} compute the $\lambda'$ correction for the
tree-point function without taking the mixing into account.  Their
result is given for $n_1=0$ (see their eq.34) which exactly
matches the one obtained from the formula (\ref{extremal}).


\section{Conclusion \label{cncl}}
Let us collect here the results of the calculations:
\begin{enumerate}
\item The extremal (two-BMN, one-BPS) correlator from
integrability (the Gromov--Vieira formalism) fully coincides at
NLO with the purely single-trace part of the perturbative extremal
correlator.
\item The extremal correlator from the string field theory with
the Dobashi-Yoneya vertex does not match the extremal correlator
from the Gromov-Vieira integrability-assisted formalism at NLO.
\item The non-extremal (three-BMN) correlators from the string
field theory and from integrability  match in the leading order
and do not match in the subleading order.
\end{enumerate}

Statement (1) has nothing remarkable in it; it merely says that
apparently no obvious mistakes have been done in the calculation
and ensures that the one-loop results at weak coupling (the
perturbative field theory and the integrability-assisted
computation) are equally adequate in describing the weakly-coupled
limit for single trace operators.

The formalism of Gromov and  Vieira does not capture the
contributions from non-single trace operators and thus cannot take
the effects related to the operator redefinition into account
which become relevant in  the case of the extremal correlator. One
should note, as had been pointed out already by Beisert et
al.\mycite{Beisert:2002bb} that the single-to-double trace mixing
matrix might also influence the $\lambda'$-order  correlator by
the terms of $\lambda'^2$-order in the mixing matrix, therefore
the $\lambda'$-terms in the weakly-coupled limit must be
considered to be reliable only after the next-next-to-leading
order of mixing matrix has been elaborated.

This means that the statement (2) potentially contains an
interplay of the effects related to this operator redefinition and
to some fundamental mismatch between the weakly and strongly
coupled theories.  Thus one should rather analyze the mismatch (3)
instead to which operator redefinition does not contribute.

What is the possible origin of this mismatch? The first guess
would be to invoke the order-of-limits argument that we have
discussed in the Introduction. This would be the most natural
explanation, yet we have seen in the story with the two-point
particle spectra that some discrepancies originally explained via
the different order of the limits had eventually found a more
physical explanation.

Furthermore, one could try to argue that there is still some
non-traced error or typo in the next-leading-order formula by
Gromov and Vieira, existence of which is not absolutely excluded,
since very few analytic calculations have been implemented using
it so far.

However, there could exist in principle a more fundamental
mismatch between the strongly and weakly coupled sectors at
next-to-leading order. This would certainly be the most intriguing
scenario, since it would challenge our current understanding of
the AdS/CFT duality for the three-point correlation function
sector.

To choose between these logical alternatives, we hope for more
tests to be performed in the nearest future, in particular those
extending beyond the $SU(2)$ sector, taking into account fermions
or considering short operators and going beyond the
Frolov-Tseytlin limit.

\section{Acknowledgements}
We would like to thank Gianluca Grignani who collaborated at the
initial stage of this project. We thank Pedro Vieira for his
valuable comments and remarks and Kolya Gromov for enlightening
correspondence. The work of W.S. is partially supported  by a Marina Solvay fellowship, by IISN - Belgium (conventions 4.4511.06 and 4.4514.08), by the ``Communaut\'e Fran\c{c}aise de Belgique" through the ARC program and by the ERC through the ``SyDuGraM" Advanced Grant. The  work of
Andrey Zayakin is funded in part by the Spanish grant
FPA2011-22594,  by Xunta de Galicia (Conseller{\'i}a de
Educaci\'on, grant INCITE09 206 121 PR and grant
PGIDIT10PXIB206075PR),  by the Consolider-Ingenio 2010 Programme
CPAN (CSD2007-00042), by FEDER, and is supported in part by the
Ministry of Education and Science of the Russian Federation under
contract 14.740.11.0081, NSh 3349.2012.2, the RFBR grants
10-01-00836 and 10-02-01483. W.S. would like to thank the physics
departments of the Universities of Perugia and Santiago de
Compostela, and A.Z. would like to thank Universit\'e Libre de
Bruxelles for the hospitality while this work was
in progress.

\appendix
\section{Theta-morphism}\label{thetamorphism}
Up to the ${\cal{O}}(g^2)$-order the $\theta$-morphism is given by
\begin{equation}
\Big(\!\!\!\Big(f(\theta)\Big)\!\!\!\Big
)=f\Bigg|_{\theta_i\rightarrow 0}+\frac{g^2}{2}\sum_{i=1}^L{\cal
D}_i^2 f\Bigg|_{\theta_i\rightarrow 0}+{\cal O}(g^4) \ ,
\end{equation}
where
\begin{equation}
{\cal D}_i=\frac{\partial}{
\partial\theta_{i}}-\frac{\partial}{\partial\theta_{i+1}}, \qquad
{\rm{with}} \qquad \theta_{L+1}=\theta_1\nonumber \ .
\end{equation}
It satisfies
\begin{eqnarray}
\Big(\!\!\!\Big(f(\theta)g(\theta)\Big)\!\!\!\Big)_{\theta}=\Big(\!\!\!\Big(f(\theta)\Big)\!\!\!\Big)_{\theta}\Big(\!\!\!\Big(g(\theta)\Big)\!\!\!\Big)_\theta+g^2\sum_{i=1}^{L}{\cal D}_if\ {\cal D}_i g \ .
\end{eqnarray}
If one of the functions is symmetric
\begin{equation}
\Big(\!\!\!\Big(f_{\rm{sym}}(\theta)g(\theta)\Big)\!\!\!\Big)_{\theta}=\Big(\!\!\!\Big(f_{\rm{sym}}(\theta)\Big)\!\!\!\Big)_{\theta}\Big(\!\!\!\Big(g(\theta)\Big)\!\!\!\Big)_\theta \ .
\end{equation}
The property which relates it to the Zhukovsky variable
\begin{equation}
\Big(\!\!\!\Big(
\sum_{i=1}^L \log\frac{u-\theta^i+i/2}{u-\theta^i-i/2}
\Big)\!\!\!\Big)_{\theta}= L\log\frac{x(u+i/2)}{x(u-i/2)}+{\cal O}(g^4) \ , \qquad x(u)=\frac{u}{g}-\frac{g}{u}+{\cal O}(g^3) \ .
\end{equation}
\section{Details for the integrability-assisted computation of the three BMN correlator}
\label{details}
In this appendix we list all  the intermediate computational steps necessary for computing the three point function of  three BMN operators of lengths $$L_1=r J+4 , \ L_2= J+2,\ L_3= J(1-r)+2$$ and rapidities
\begin{eqnarray}
{\cal O}_1&: & u_1,-u_1,u_3,-u_3\nonumber\\
{\cal O}_2&: & v,-v\nonumber\\
{\cal O}_3&: & w,-w
\end{eqnarray}
which up to the ${\cal{O}}(g^2)$-order are given by
\begin{eqnarray}
u_1&=&\frac{Jr+3}{2\pi n_1}+g^2\frac{4\pi n_1}{Jr+3}\nonumber\\
u_3&=&\frac{Jr+3}{2\pi n_4}+g^2\frac{4\pi n_4}{Jr+3}\nonumber\\
v&=&\frac{J-1}{2\pi n_2}+g^2\frac{4\pi n_2}{J-1}\nonumber\\
w&=&\frac{J(1-r)+1}{2\pi n_3}+g^2\frac{4\pi n_3}{J(1-r)+1}
\end{eqnarray}

For the computation below we also need to know the momenta of the third operator which is up to the ${\cal{O}}(g^2)$-order
\begin{equation}
p_1^{(3)}=-p_2^{(3)}=\frac{2\pi n_3}{J(-1+r)}\left(
1-\frac{g^2}{J^2}\frac{8\pi^4n_3^4}{J^2(-1+r)^4}
\right)
\end{equation}
\begin{itemize}
\item
The norms with the $\theta$-morphism are given by
\begin{eqnarray}
\Big(\!\!\!\Big(\langle\theta ;{\bf{u}}|\theta ;{\bf{u}}\rangle\Big)\!\!\!\Big)_{\theta}&=&\frac{(2\pi n_1)^8}{J^4r^4}\left(1+\frac{g^2}{J^2} \frac{8(2\pi n_1)^2
}{r^2}\right)\, ,\nonumber\\
\Big(\!\!\!\Big(\langle\theta ;{\bf{v}}|\theta ;{\bf{v}}\rangle\Big)\!\!\!\Big)_{\theta}&=&\frac{(2\pi n_2)^4}{J^2}\left(
1+\frac{g^2}{J^2}4(2\pi n_2)^2
\right)\, ,\nonumber\\
\Big(\!\!\!\Big(\langle\theta ;{\bf{w}}|\theta ;{\bf{w}}\rangle\Big)\!\!\!\Big)_{\theta}&=&\frac{(2\pi n_3)^4}{J^2(1-r)^2}\left(
1+\frac{g^2}{J^2}\frac{4(2\pi n_3)^2}{(1-r)^2}\right) \, .
\end{eqnarray}

The denominator in the final expression (\ref{GVfinal}) is given by
\begin{eqnarray}
\Bigg(\Big(\!\!\!\Big(\langle\theta ;{{\bf{u}}}|\theta ;{\bf{u}}\rangle\Big)\!\!\!\Big)_{\theta}\Big(\!\!\!\Big(\langle\theta ;{\bf{v}}|\theta ;{\bf{v}}\rangle\Big)\!\!\!\Big)_{\theta}\Big(\!\!\!\Big(\langle\theta ;{\bf{w}}|\theta ;{\bf{w}}\rangle\Big)\!\!\!\Big)_{\theta}\Bigg)^{-1/2}& = &
(2\pi)^8\frac{n_1^2 n_2^2n_3^2n_4^2}{(Jr)^2[J(1-r)] J}
\times\nonumber\\ &&\! \! \! \! \! \! \! \! \! \! \! \! \! \! \! \! \! \! \! \! \! \! \! \! \! \! \! \! \! \! \! \! \! \! \! \! \! \! \! \! \! \! \! \! \! \! \! \! \! \! \! \! \! \! \! \! \! \! \! \! \! \! \! \! \times
\left(1+\frac{g^2}{J^2}8\pi^2\left(
\frac{n_1^2+n_4^2}{r^2}+n_2^2+\frac{n_3^2}{(-1+r)^2}
\right)\right)\, .
\end{eqnarray}

\item{The normalization coefficient}
 $\sqrt{L_1 L_2 L_3}$ in the large $J$-limit is given by
\begin{equation}
\sqrt{L_1 L_2 L_3}=\sqrt{Jr \ J \  (1-r)J} \, .
\end{equation}
\item
$\tilde {\cal A}$ for the operator with two magnons and rapidities $w,-w$ becomes
\begin{equation}
\tilde{\cal{A}}_{N_3}({\bf{w}})=\Big(1-\mathfrak{f}(p_1,p_2)e^{2ip_2}-\mathfrak{f}(p_2,p_1)e^{2ip_1}+e^{2i(p_1+p_2}   \Big)\nonumber \ ,
\end{equation}
\begin{equation}
{\mathfrak{f}}(p_i,p_j)=\left(1+\frac{i}{w(p_i)-w(p_j)}\right)\left(1+\frac{g^2}{(w(p_i)^2+\frac{1}{4})(w(p_j)^2+\frac{1}{4})}\right)\nonumber
\end{equation}
which combined with $\left(1-g^2\frac{\Gamma_{\bf w}}{2}\right)$ gives
\begin{equation}
\left(1-g^2\frac{\Gamma_{\bf w}}{2}\right)\tilde {\cal A}_{N_3}=\frac{8\pi^2n_3^2}{J^2(-1+r)^2}+{\cal O}(g^4) \, .
\end{equation}

\item Computationally, the most complicated expression is $\Big(\!\!\!\Big(\langle \theta^{1};{\bf{u}} |{\cal O}_3| \theta^{(2)};{\bf {v}}\rangle\Big)\!\!\!\Big)$.
We use the property of the theta-morphism applied to a product of factors. Then we Taylor expand the determinant expression up to cubic order in $\theta$ and execute the theta-morphism up to ${\cal{O}}(g^2)$-order.
\begin{eqnarray}
\Big(\!\!\!\Big(\langle \theta^{1};{\bf{u}} |{\cal O}_3| \theta^{(2)};{\bf {v}}\rangle\Big)\!\!\!\Big)_{\theta^{(1)}} &=&\Big(\!\!\!\Big({\rm {prefactor}}\times \det \Big([ G_{mn}]\oplus [ F_{nm}]\Big) \Big)\!\!\!\Big)_{\theta^{(1)}}\nonumber\\
&=&(\theta^{(1)}_1-\theta^{(1)}_2)\Big(\!\!\!\Big({\rm {prefactor}} \Big)\!\!\!\Big)_{\theta^{(1)}} \Big(
D^{0,0}_{g_0}+g^2 (D^{0,0}_{g_2}+ 3D^{0,2}_{g_0}-D^{1,1}_{g_0})
\Big)\nonumber\\
&&+{\rm{\ cross\ term}}
\end{eqnarray}
with
\begin{equation}
D^{p,q}=\det \Bigg(
G_{nm}\Big|_{\theta_a=0} \ \oplus \Big(\Phi_n^p \ \Phi_n^{q+1}\Big)
\Bigg)
\ , \qquad \Phi^p=\frac{1}{p!}\left(-\frac{\partial}{\partial u_n}\right)^p\frac{1}{u_n^2+1/4}\, .\nonumber
\end{equation}
$n$ goes in our case from 1 to $N_1$, which makes $\Big(\Phi_n^p \ \Phi_n^{q+1}\Big)$ two by four matrix.

The cross term which comes from $g^2 \sum_i {\cal D}_i ({\rm{prefactor}})\times {{\cal D}}_i (\rm{determinant})$ includes a factor of $D^{0,1}$ which is zero.

\begin{equation}
(\theta^{(1)}_1-\theta^{(1)}_2)\Big(\!\!\!\Big({\rm {prefactor}} \Big)\!\!\!\Big)_{\theta^{(1)}}=\frac{\pi^3 r^2}{J^3}\frac{n_1n_4 n_2^5}{(n_1^2-n_4^2)^2}\left(
1+\frac{g^2}{J^2}\frac{8\pi^2}{r^2}(n_1^2+n_4^2-3n_2^2 r^2)
\right)\, ,
\end{equation}

\begin{equation}
\Big(\!\!\!\Big(\langle\theta^{(1)} ;{\bf{u}}| \hat{\cal{O}}_3 |\theta^{(1)} ;{\bf{v}}\rangle\Big)\!\!\!\Big)_{\theta^(1)}\Bigg|_{n_4\rightarrow n_1}
=-\frac{32(2\pi)^4}{J^4 r^2}\frac{n_1^4 n_2^2(n_1^2+3n_2^2 r^2)}{(n_1-n_2^2 r^2)^2}\left(
1+\frac{g^2}{J^2}\frac{(2\pi)^2}{ r^2}\frac{3n_1^4+17 n_1^2 n_2^2 r^2}{n_1^2+3 n_2^2 r^2}
\right) \, .
\end{equation}

\end{itemize}

Combining all the intermediate results together gives
\begin{eqnarray}\label{GVfinal}
C_{GV,3BMN}&=&\frac{16\sqrt{r}\sin^2(\pi n_2 r)}{\sqrt{J(1-r)}\pi^2}\ \frac{n_1^2+3n_2^2 r^2}{(n_1^2-n_2^2 r^2)^2}
\left(
1+\frac{g^2}{J^2} 4\pi^2\left(
-\frac{2n_3^2}{(1-r)^2}-\frac{n_1^2}{r^2}-2n_2^2+\frac{8n_1^2n_2^2}{n_1^2+3n_2^2r^2}
\right)
\right)\, .\nonumber\\
\end{eqnarray}



\providecommand{\href}[2]{#2}\begingroup\raggedright\endgroup

\end{document}